\documentclass[11pt,twoside]{article}
\usepackage{epsfig}
\usepackage{here}
\usepackage{amsmath}
   
\begin{document}

\newcommand{\hppee}{\mbox{$\et\rightarrow\pi^+\pi^-e^+e^-$}}
\newcommand{\hppg}{\mbox{$\et\rightarrow \pi^-\pi^+\gamma$}}
\newcommand{\et}{\mbox{$\eta$}}
\newcommand{\etp}{\mbox{$\eta^{\prime}$}}
\newcommand{\hppp}{\mbox{$\et\rightarrow \pi^-\pi^+\pi^0$}}
\newcommand{\ppeta}{\mbox{$pp\rightarrow pp\et$}}
\newcommand{\pneta}{\mbox{$pn\rightarrow pn\et$}}
\newcommand{\pnetap}{\mbox{$pn\rightarrow pn\et^{\prime}$}}

\begin{center}{\bf\Large Quasi-free production  
   of the $\boldsymbol{\eta}$ and  $\boldsymbol{\eta}^{\boldsymbol{\prime}}$ mesons in nucleon-nucleon collisions} 
\end{center}
\vspace{0.5cm}
\begin{center}
   J. Stepaniak $^{a}$, H.  Cal\'en $^{b}$ \\ 
   {\small \em 
         $^a$Soltan Institute for Nuclear Studies, Warsaw, Poland\\
         $^b$The Svedberg Laboratory, Uppsala, Sweden
   }
\end{center}
\vspace{0.5cm}
\begin{center}
 \parbox{0.9\textwidth}{
  \small{
    {\bf Abstract:}\
     Some aspects of the quasi-free production of mesons in the near threshold 
     region are discussed.
     A method of data analysis used  by WASA/PROMICE Collaboration to extract
     the excitation function  for \et\ production on neutron target
     is briefly presented.     Feasibility of similar study with
     WASA at COSY is discussed.

     } 
      
 }
\end{center}
\vspace{0.5cm}
\section{Introduction}
It is rather obvious, that the Fermi motion changes the kinematics of the
interaction on the nucleon in nuclei, as compared to that on the 
target nucleon at rest, but the importance of the effect in the near threshold region is not always sufficiently appreciated, especially for the cross section
measurements. The control of the energy available in the quasi free nucleon-nucleon scattering on the event by event basis is of importance.
We will discuss here the method of     extraction of the
cross sections on protons and neutrons from proton-deuteron interaction
applied by the WASA Collaboration to the \et\ meson production \cite{pneta}
as well as the possibility of using the same method for the \etp\ production
with WASA at COSY.

The energy available in the center of mass of the nucleon-nucleon system
is:
\begin{equation}
s=(E_b+E_t)^2 - (\vec{p}_b+\vec{p}_t)^2 =(E_b+E_t)^2 - p_b^2 -p_t^2 - 2 p_b p_t  cos(\theta)
\end{equation}
where $\vec{p}_t$ is the target momentum and $\vec{p}_t=-\vec{p}_{Fermi}$.

The product $p_b p_t$ is not small at high energies of the beam. It is multiplied by the cosine of the angle of the target momentum with respect to the beam. 
So it is mainly the longitudinal Fermi momentum component, that influence the energy available in the reaction.

\begin{figure}[H]
\parbox{0.4\textwidth}
   {\epsfig{file=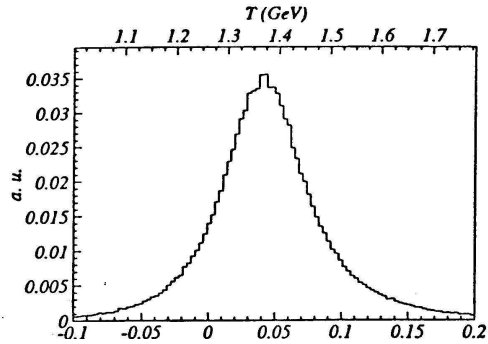,width=0.6\textwidth}}
\hfill
\parbox{0.4\textwidth}
  {\caption{\label{Q_fig} \small Distribution of the calculated CM excess energy in the nucleon-nucleon system for the 1360 MeV proton beam. 
   }}
\end{figure}

 The importance of the Fermi momentum effect is quite
frequently underestimated, especially for high energy beams.  In the near threshold region, where the cross section is a very steep function of  energy it can dramatically influence the observed cross section.
 The most probable value of the Fermi momentum in the deuteron is small, equal to  about 50 MeV/c. It seems  
to be negligible as compared to a few GeV beam energy. In reality e.g. at 1360MeV beam energy the excess energy in the nucleon-nucleon CM system can be changed by 43.5 MeV  when 50MeV/c target is moving upstream the beam . At 2500 MeV the change is as large as 58.7 MeV. Such a shift can change the    
\ppeta\ cross section near threshold by an order of magnitude. The distribution of the excess of energy in the CM system for the eta meson production in \ppeta\ reaction at 1360 MeV incident energy is shown in figure~\ref{Q_fig}.
\vspace{-0.4cm}
\vspace{-1.5cm}
\begin{figure}[H]
\parbox{0.4\textwidth}
   {\epsfig{file=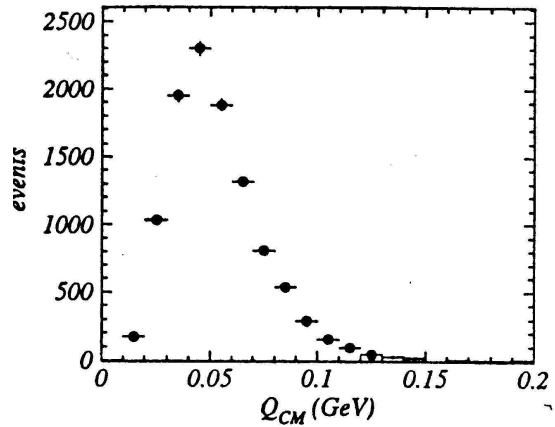,width=0.6\textwidth}}
\hfill
\parbox{0.4\textwidth}
  {\caption{\label{Q_exp} \small Distribution of the measured $Q_{CM}$ in the nucleon-nucleon system for the 1360 MeV proton beam. 
   }}
\end{figure}

\begin{figure}[H]
\parbox{0.5\textwidth}
   {\epsfig{file=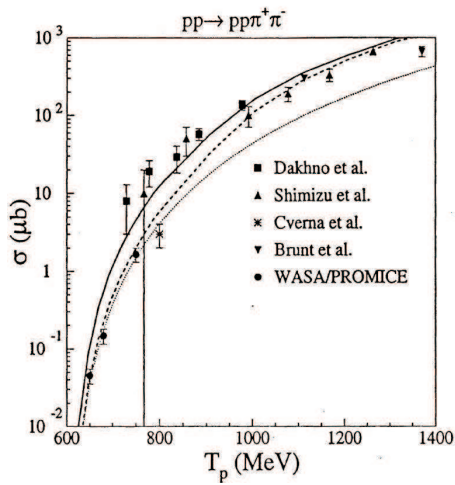,width=0.5\textwidth}}
\hfill
\parbox{0.5\textwidth}
  {\caption{\label{pipi_bmp} \small Cross section for
  the reaction $pp\rightarrow pp\pi^+\pi^-$ as a function of beam energy.
    Solid squares are from an experiment
    performed on deuteron target. The lines show theoretical predictions.  
   }}
\end{figure}

 Measured distribution of the $Q_{CM}$ for the $p+n\rightarrow d+\et$ quasi-free reaction is shown in figure~\ref{Q_exp}. The distribution is not symmetric and the  average value
of the  cross section does not correspond to the nominal beam energy
 because of the convolution with the energy dependent cross section.
 One can easily find
 examples of the papers in which  hadron-nucleon cross sections were extracted from the quasi-free interactions on the deuteron without any correction for the effective energy in the nucleon-nucleon system.
 It leads to an overestimation of the cross section measured in a quasi-free interaction even on the so loosely bound nucleus as a  deuteron.
Let us take as an example the cross section for the two-pion production in the threshold region  (see figure~\ref{pipi_bmp} ).  The points from the experiment performed on the deuteron (full squares in the fig.~\ref{pipi_bmp} are substantially higher than the points from the experiments performed on free protons. The authors claimed that `` The Fermi motion 
in the deuteron leads to a small enlargement of the beam momentum distribution only''
and their measured cross sections were attached to the nominal energies of the beam. In fact the distribution of the energy in nucleon-nucleon CMS in not
symmetric, when folded with the cross section and the average corresponding
  beam energy should be shifted to the right quite substantially.

In the naive understanding of the spectator model the spectator nucleons have
to be emitted isotropically with  momentum distribution following 
the squared wave function of nucleons in the deuteron. It is usually wrong near the reaction
threshold.
 The forward-backward asymmetry of spectators can be
very pronounced and the tail of high momentum enhanced. It is due
to the effective Q value dependence on the direction of the Fermi momentum
vector.   

\section{Cross section extraction method for quasi-free \et\ production}
In the study performed by WASA/PROMICE collaboration \cite{pneta,Stina}
 validity of the impulse approximation was assumed. All the observed events 
were classified into the following classes of reactions:
\begin{enumerate}
\item
$p+d \rightarrow d+\eta (p_s)$
\item
$p+d \rightarrow p+p+\eta (n_s)$
\item
$p+d \rightarrow p+n+\eta (p_s)$
\item
$p+d \rightarrow ^3\!\!He + \eta$
\end{enumerate}
where the s subscript denotes a slow spectator nucleon for the dominant 
quasi-free reaction.
It is impossible to judge which nucleon was active in the $\eta$ production 
before the $^3He$ formation. Therefore we neglected the small fraction
of the events with helium nucleus in the final state. 

The possible double
scattering and screening effects are expected to be on the level of few 
percent. We neglected it as well. It defines the accuracy of the method.

All the events with the two photons invariant  mass consistent with the \et\ mass were divided into two classes according to the event topology:
the events with two charged in the forward part of the detector
were attributed to the reaction 2, those with one particle were candidates
for the reaction 1 or 3. The distinction between reaction 1 and 3 was
possible because of the almost two-body kinematics in the first case.
The $Q_{CM}$  value was calculated from the measured quantities
if only one particle was not observed. In the case of the reaction 3
the transverse component of the target nucleon momentum was neglected
to reconstruct the energy excess $Q_{CM}$. The resolution in $Q_{CM}$ was
found to be better than 10 MeV even in this case. It was checked using
the Monte Carlo simulation of the process.     

\begin{figure}[H]
\epsfig{file=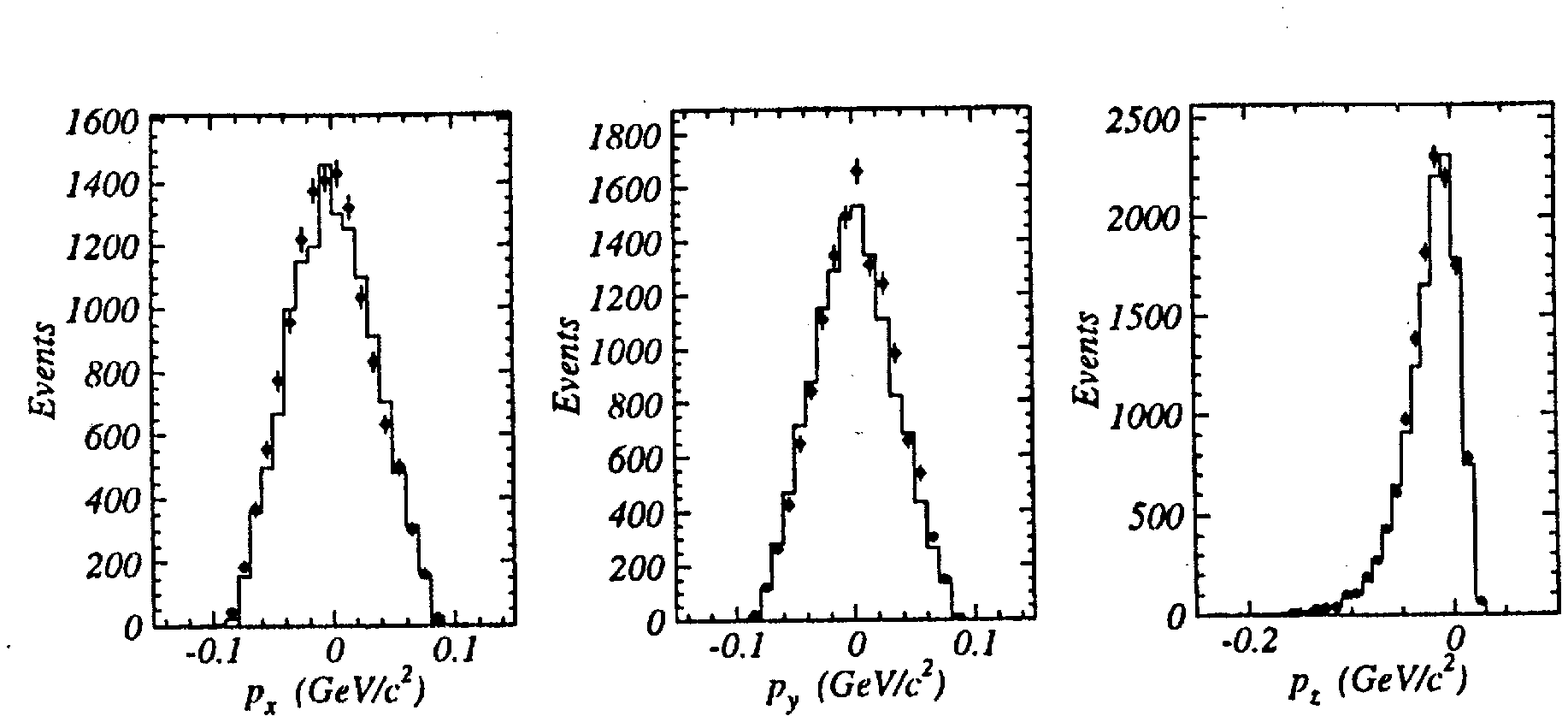,width=0.9\textwidth}
\caption{\label{pxpypz} \small Distributions of the momentum component of the target neutron (opposite to the spectator nucleon) from $pn\rightarrow d\eta$ reaction at 1360 MeV; histograms - Monte Carlo, points - the experimental data. }
\end{figure}

In the Monte Carlo generation of the quasi-free pN reactions one should assure the energy conservation. Since spectator is on shell it requires:\\
$E_{target}=m_d - E_{s} = m_d - \sqrt{m_s^2 + p_s^2}$,
where the subscript s stands for ``spectator''.
The reconstructed momentum of target neutron for the reaction $pn\rightarrow d\eta$ was compared with the result of Monte Carlo generation. Monte Carlo 
events were weighted with the measured cross section.
The agreement between MC simulated  experimental distributions was observed
for several variables. As an example the distributions of the momentum
component of the target neutron is shown in figure ~\ref{pxpypz}.
It confirmed validity of the assumed model of quasi-free interaction as well as the generation method. 

\begin{figure}[H]
%\parbox{0.5\textwidth}
{\epsfig{file=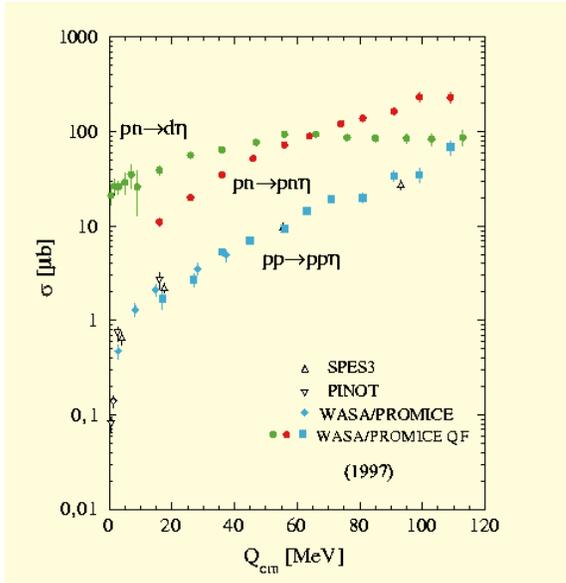,width=0.6\textwidth}}
\hfill
%\parbox{0.5\textwidth}
    {\caption{\label{eta_qf} \small Cross sections for \et\ meson production as a function of the CM excess of energy. }}
\end{figure}

The result  of the cross section measurements  are presented in figure~\ref{eta_qf}. The points for \ppeta\ reaction from the quasi-free interaction are 
in agreement (within errors) with points from the reaction on free protons
It is an additional argument that the impulse approximation can be used to describe the \et meson production on deuteron at the energy considered and the excitation function can be extracted from the measurement of the quasi-free interactions from the experiment performed at one incident energy.

\section{Quasi-free production of \etp\ with WASA at COSY}
The feasibility study of the quasi-free creation of the \etp\ meson in the reaction \pneta\   was presented by P.~Moskal \cite{Moskal} at COSY-11 Collaboration Meeting 2001. The idea of the experiment was based on the simultaneous measurement of the momenta of spectator proton and the neutron in COSY-11 setup.
With WASA at COSY the alternative method : measurement of the \et\ or \etp\ decay products will be possible. The table \ref{tabelka} summarize
the acceptances and neutral decay channels branching ratios at the selected
beam energies \cite{Calen}.
\begin{table}[H]
    \caption{\label{tabelka}\small Comparison of BR and acceptances for the \et\ and \etp\ production
         in the pp interactions at energy excess $Q_{cm}=40$ MeV.}
       
          \begin{tabular}{|l|l|l|}
          \hline
            & \et\ $E_{inc}=$1360 MeV & \etp\  $E_{inc}=2550$ MeV \\ \hline \hline
         Total cross section & 5$\mu$b & 0.25 $\mu$b \\ \hline
         Branching ratio:& & \\
         $\gamma\gamma$ & 0.39 & 0.02 \\ 
         $\pi^0\pi^0\pi^0$ & 0.325 & 0.002 \\
         $\pi^0\pi^0\eta$ & -   & 0.21 \\ \hline 
         Acceptance:& & \\
          $\gamma\gamma$ & 66\% & 52\% \\ 
          $\pi^0\pi^0\pi^0$ & 12\% & 6\% \\
          $\pi^0\pi^0\eta$ & -  & 4.5 \\ \hline \hline
    \end{tabular}
\end{table}

According to our experience at CELSIUS/WASA the neutral decay channels can provide more selective trigger and background condition than the charged ones.
The contribution from prompt three $\pi^0$ production is much smaller then
in the case of channels containing charged pions. 

\section{Summary}
 
The Collaboration exploited the spread of the energy observed in nucleon
nucleon quasi-free interactions at one incident proton energy (1360 MeV) 
to measure
the energy  dependence of the cross section for the \et\ meson production.
In particular it permitted to study the production in proton-neutron scattering,
which was not available before.

To achieve this the Q value was measured on an event by event basis in the
nucleon - active nucleon system.
It became possible due to the WASA set-up ability to measure the momenta of all the active  reaction products including  photons from the meson decay.

\end{document}